\documentclass[aps,prb,twocolumn,groupedaddress,amsmath,letterpaper]{revtex4}

\usepackage{graphicx}
\usepackage{array}

\begin{document}

\title{Electro-magnetostatic homogenization of bianisotropic metamaterials.}

\author{Chris Fietz}
\email[Email: ]{fietz.chris@gmail.com}
\affiliation{Ames Laboratory and Department of Physics and Astronomy, Iowa State University, Ames, Iowa 50011, USA}

\begin{abstract}
We apply the method of asymptotic homogenization to metamaterials with microscopically bianisotropic inclusions to calculate a full set of constitutive parameters in the long wavelength limit.  Two different implementations of electromagnetic asymptotic homogenization are presented.  We test the homogenization procedure on two different metamaterial examples.  Finally, the analytical solution for long wavelength homogenization of a one dimensional metamaterial with microscopically bi-isotropic inclusions is derived.
\end{abstract}

\pacs{}

\maketitle


\section{Introduction}\label{Intro}

The foundational claim of the metamaterial research community is the ability to engineer artificial materials with deliberate electromagnetic responses.  However, despite considerable developments in the metamaterial field over the last 10 plus years, the question of how to quantify the electromagnetic response of a metamaterial is still very much an open one.  The electromagnetic response of any material is quantified by the so called constitutive parameters of the materials (permittivity, permeability, etc.)  The determination of these constitutive parameters in the long wavelength limit is the subject of this paper.

One of the earliest attempts to assign values to the constitutive parameters of a metamaterial was the method sometimes referred to as as S-parameter retrieval~\cite{Soukoulis_02}.  This method lacks generality in that it assumes that the metamaterial in question is isotropic and has no electromagneto coupling, though it is recognized that it can be used to characterize anisotropic materials with diagonal constitutive tensors.  It has since been generalized to accommodate bi-anisotropic metamaterials obeying certain symmetry requirements~\cite{Li_09} as well as reciprocal bi-isotropic (isotropic chiral) metamaterials~\cite{Wang_09}.  In addition, there have been many other modifications of the original S-parameter retrieval method which are less useful, either because they are trivial or because they use the retrieval method in ways that are inappropriate considering the assumptions the original method is based on.  In addition to the fact that S-parameter retrieval is only applicable to highly symmetric metamaterials, it should also be noted that this method assumes no spatial dispersion in the material it is applied to.  This simplification affects many other assumptions implicit in the retrieval method regarding the symmetry of the constitutive parameters and the boundary conditions of the material, limiting the validity of the S-parameter retrieval method, as well as introducing apparently non-physical characteristics to the retrieved constitutive parameters due to the presence of spatial dispersion~\cite{Koschny_03}.  Despite these limitations, S-parameter retrieval is today the most widely used method for assigning values to the constitutive parameters of metamaterials.

  In addition to S-parameter retrieval, there have been several attempts at developing a more general homogenization theory for metamaterials~\cite{Smith_06,Silveirinha_07,Li_07,Simovski_07_Bloch,Fietz_10b,Alu_11,Pors_11}.  There has been limited success with these methods but none can be considered a general theory for metamaterial homogenization.  One trait that all of these methods have in common is that none of them are related to the classical method of homogenization known as asymptotic homogenization.
  
  Asymptotic homogenization, sometimes referred to as classical homogenization, is a old and established method for homogenizing differential equations that is limited to the long wavelength limit.  For an introduction to the asymptotic homogenization method see Refs.~\onlinecite{Brezis_94,Hornung_96,Bensoussan_78}.  The fact that asymptotic homogenization only works in the long wavelength limit is the main reason that it has been largely ignored by the metamaterial theory community.  Many of the more interesting metamaterial phenomenon, negative index of refraction for example, involve resonances which difficult to characterize using asymptotic homogenization.  However, one advantage of asymptotic homogenization is that it can characterize very asymmetric and therefore anisotropic materials, something many other homogenization procedures cannot do.  There have been a number of attempts to apply this method to metamaterials~\cite{Banks_05,Ouchetto_06,Cao_10}.  In addition there is a similar homogenization method that involves placing a metamaterial between two metal plates and calculating the macroscopic permittivity by treating the metamaterial as a capacitor~\cite{Brosseau_01,Urzhumov_07}.  This method is less general than asymptotic homogenization in that it requires certain symmetries in the metamaterial to be performed correctly.  However, within this limitation the method is mathematically equivalent to asymptotic homogenization.  In  this paper we apply the asymptotic theory of homogenization to metamaterials with microscopically bianisotropic inclusions.  Our use of the method is distinguished from many of the previously mentioned references in that we work with potential fields, as opposed to the electric and magnetic fields.

\section{Asymptotic homogenization}\label{Sec_2}

\subsection{The electromagnetic four-potential}\label{four_po}

The normal implementation of asymptotic homogenization method is typically applied to a potential field.  One example of this is the homogenization of electrostatic permittivity tensors using the electrostatic potential.  Other examples include the homogenization of a stiffness tensor using the displacement vector as the potential field, the homogenization of thermal conductivity using the temperature as the potential field, and the homogenization of a permeability tensor using density as the potential field.  Our first implementation of the asymptotic homogenization will use the electromagnetic four-potential, consisting of the electrostatic scalar potential $\mathrm{A}_0$ and the electromagnetic vector potential $\textbf{A}$.  Following the standard method of asymptotic expansion, we assume that the four-potential can be represented as an expansion in terms of the unitless small parameter $\alpha$

\begin{equation}\label{expan}
\begin{array}{rl}
\mathrm{A}_0^{\alpha}(\textbf{x}) = & \mathrm{A}_0^0(\textbf{x},\textbf{y}) + \alpha \mathrm{A}_0^1(\textbf{x},\textbf{y}) + \alpha^2 \mathrm{A}_0^2(\textbf{x},\textbf{y}) + \cdots\ , \\[5pt]
\textbf{A}^{\alpha}(\textbf{x}) = & \textbf{A}^0(\textbf{x},\textbf{y}) + \alpha \textbf{A}^1(\textbf{x},\textbf{y}) + \alpha^2 \textbf{A}^2(\textbf{x},\textbf{y}) + \cdots\ .
\end{array}
\end{equation}  

\noindent Here $\textbf{y}=\textbf{x}/\alpha$ is a vector coordinate describing the rapidly changing values of the four-potential inside the unit cell and, $\textbf{x}$ is a vector coordinate describing the slowly changing values of the four-potential on a larger spatial scale.  The potential fields are periodic with respect to $\textbf{y}$, with periodicity given by the lattice constants of the unit cell.  There is no requirement that the unit cell be cubic or tetragonal, only that it is periodic.  The electric field and magnetic flux density are defined as

\begin{equation}\label{EB_def}
\begin{array}{rl}
\textbf{E} = & -\nabla \mathrm{A}_0^{\alpha}, \\ \\
\textbf{B} = & \nabla\times\textbf{A}^{\alpha}.
\end{array}
\end{equation}

\noindent  Here and throughout the rest of this paper we use Heaviside-Lorentz units.  In Eq.~(\ref{EB_def}) we have assumed that the fields are static.  These fields are related to the electric displacement and the magnetic field by the microscopic constitutive relations

\begin{equation}\label{Con1}
\left(\!\!\!\begin{array}{c}
\textbf{D} \\ \textbf{H}
\end{array}\!\!\!\right) = 
\underbrace{\left(\!\!\begin{array}{cc}
\hat{p} & \hat{l} \\ \hat{m} & \hat{q}
\end{array}\right)}_{\hat{K}}\cdot
\left(\!\!\begin{array}{c}
\textbf{E} \\ \textbf{B}
\end{array}\!\!\!\right).
\end{equation}

\noindent Here the constitutive matrix $\hat{K}$ consists of $3\times 3$ tensors ($\hat{p}$ and $\hat{q}$) and pseudotensors ($\hat{l}$ and $\hat{m}$). The microscopic constitutive parameters also vary according to both the large and small scale variables $\hat{K}=\hat{K}(\textbf{x},\textbf{y})$ and are periodic with respect to $\textbf{y}$.  These constitutive relations differ from the more typical constitutive relations

\begin{equation}\label{Con2}
\left(\!\!\!\begin{array}{c}
\textbf{D} \\ \textbf{B}
\end{array}\!\!\!\right) = 
\underbrace{\left(\!\!\begin{array}{cc}
\hat{\epsilon} & \hat{\xi} \\ \hat{\zeta} & \hat{\mu}
\end{array}\right)}_{\hat{C}}\cdot
\left(\!\!\begin{array}{c}
\textbf{E} \\ \textbf{H}
\end{array}\!\!\!\right),
\end{equation}

\noindent the relationship between the different microscopic constitutive parameters of Eqs.~(\ref{Con1}) and~(\ref{Con2}) being

\begin{equation}
\begin{array}{rl}
\hat{p} = & \hat{\epsilon}-\hat{\xi}\cdot\hat{\mu}^{-1}\cdot\hat{\zeta}, \\[5pt]
\hat{l} = & \hat{\xi}\cdot\hat{\mu}^{-1}, \\[5pt]
\hat{m} = & -\hat{\mu}^{-1}\cdot\hat{\zeta}, \\[5pt]
\hat{q} = & \hat{\mu}^{-1}.
\end{array}
\end{equation}

  The differential operator acting on the four-potential in Eq.~(\ref{EB_def}) is simply a four dimensional exterior derivative (with the time derivative equal to zero) applied to a four-vector.  In the traditional asymptotic homogenization, the equation of motion is an divergence (an interior derivative) applied to the constitutive matrix times the gradient (an exterior derivative) of the scalar potential.  Similarly, our equation of motion is the interior derivative of the electric displacement and magnetic fields, which in turn are simply the constitutive matrix $\hat{K}$ times the exterior derivative of the four-potential
  
\begin{equation}\label{field_eq}
\begin{array}{c}
0 = \nabla\cdot\textbf{D} = \nabla\cdot\left[\hat{p}\cdot(-\nabla \mathrm{A}_0^{\alpha})+\hat{l}\cdot(\nabla\times\textbf{A}^{\alpha})\right], \\ \\
0 = \nabla\times\textbf{H} = \nabla\times\left[\hat{m}\cdot(-\nabla \mathrm{A}_0^{\alpha})+\hat{q}\cdot(\nabla\times\textbf{A}^{\alpha})\right].
\end{array}
\end{equation}

\noindent Again we have assumed that the fields are static.  One last detail is our choice of gauge

\begin{equation}\label{gauge}
\nabla\cdot\textbf{A}^{\alpha} = 0.
\end{equation}

\noindent  In the static regime the Lorenz gauge coincides with the Coulomb gauge.  This gauge condition must be satisfied in addition to the field equation.

As per the standard asymptotic homogenization method, we insert the field expansion of Eq.~(\ref{expan}) into the field equation Eq.~(\ref{field_eq}) and gauge condition Eq.~(\ref{gauge}), separating the resulting equations according to the order of $\alpha$.  The field equation proportional to $\alpha^{-2}$ is

\begin{equation}\label{alpha_2}
\left(\!\!\!\begin{array}{c}
\nabla_y\cdot \\ \nabla_y\times
\end{array}\!\!\!\right)
\cdot\left[\hat{K}\cdot
\left(\!\!\!\begin{array}{c}
-\nabla_y\mathrm{A}_0^0 \\ \nabla_y\times\textbf{A}^0
\end{array}\!\!\!\right)\right] = 0.
\end{equation}

\noindent  Here the operator $\nabla_y$ only acts on the $\textbf{y}$ variable of the potential field.  The $\alpha^{-1}$ order gauge condition is 

\begin{equation}
\nabla_y\cdot\textbf{A}^0 = 0.
\end{equation}

\noindent This gauge condition combined with Eq.~(\ref{alpha_2}) and the requirement of periodicity with respect to $\textbf{y}$ implies that the $0$th order fields are constant with respect to $\textbf{y}$ or $\mathrm{A}_0^0=\mathrm{A}_0^0(\textbf{x})$ and $\textbf{A}^0=\textbf{A}_0^0(\textbf{x})$.

The $\alpha^{-1}$ order field equation is

\begin{equation}\label{alpha_1}
\left(\!\!\!\begin{array}{c}
\nabla_y\cdot \\ \nabla_y\times
\end{array}\!\!\!\right)\cdot
\left[\hat{K}\cdot
\left(\!\!\!\begin{array}{c}
-\nabla_x\mathrm{A}_0^0-\nabla_y\mathrm{A}_0^1 \\ \nabla_x\times\textbf{A}^0+\nabla_y\times\textbf{A}^1
\end{array}\!\!\!\right)\right] = 0.
\end{equation}

\noindent Here we have taken advantage of the fact that $\mathrm{A}_0^0$ and $\textbf{A}^0$ are independent of $\textbf{y}$.  We solve this equation by relating the first order four-potential to the derivatives of the $0$th order four-potential

\begin{equation}\label{repre_1}
\begin{array}{rl}
\mathrm{A}_0^1(\textbf{x},\textbf{y}) = & \displaystyle\sum_{i=1}^6
\left(\!\!\!\begin{array}{c}
-\nabla_x\mathrm{A}_0^0(\textbf{x}) \\ \nabla_x\times\textbf{A}^0(\textbf{x})
\end{array}\!\!\!\right)_i
\mathrm{a}_{0i}(\textbf{y}), \\[5pt]
\textbf{A}^1(\textbf{x},\textbf{y}) = & \displaystyle\sum_{i=1}^6
\left(\!\!\!\begin{array}{c}
-\nabla_x\mathrm{A}_0^0(\textbf{x}) \\ \nabla_x\times\textbf{A}^0(\textbf{x})
\end{array}\!\!\!\right)_i
\textbf{a}_i(\textbf{y}).
\end{array}
\end{equation}

\noindent  Here the subscript $i$ on the $6\times1$ vector indicates the $i'th$ component of that vector.  We have introduced two new fields, $\mathrm{a}_{0i}(\textbf{y})$ and $\textbf{a}_i(\textbf{y})$, which are typically called \textit{correctors}.  We ensure the periodicity of $\mathrm{A}_0^1$ and $\textbf{A}^1$ by forcing $\mathrm{a}_{0i}(\textbf{y})$ and $\textbf{a}_i(\textbf{y})$ to be periodic with respect to $\textbf{y}$.  With this representation we can satisfy Eq.~(\ref{alpha_1}) by solving the so-called cell problem

\begin{equation}\label{cell_1}
\left(\!\!\!\begin{array}{c}
\nabla_y\cdot \\ \nabla_y\times
\end{array}\!\!\!\right)
\cdot\left[\hat{K}\cdot\left(\hat{\textbf{e}}_j + 
\left(\!\!\!\begin{array}{c}
-\nabla_y\mathrm{a}_{0j} \\ \nabla_y\times\textbf{a}_j
\end{array}\!\!\!\right)\right)\right] = 0.
\end{equation}

\noindent Here $\hat{\textbf{e}}_i$ is a $6\times 1$ unit vector pointing in the $i$th direction such that $\hat{\textbf{e}}_i\cdot\hat{\textbf{e}}_j=\delta_{ij}$.

The cell problem must be solved along with the $\alpha^0$ order gauge condition.

\begin{equation}
\nabla_y\cdot\textbf{A}^1 + \nabla_x\cdot\textbf{A}^0 = 0.
\end{equation}

\noindent This equation must be true in order to satisfy the original gauge condition, but since it involves two fields we have additional freedom in how we satisfy it.  The simplest choice is to use the Coulomb gauge for both fields, giving us

\begin{equation}\label{gauge1}
\nabla_y\cdot\textbf{A}^1 = 0,
\end{equation}

\noindent and

\begin{equation}\label{gauge0}
\nabla_x\cdot\textbf{A}^0 = 0.
\end{equation}

\noindent Equation~(\ref{gauge0}) is a gauge condition for the macroscopic potential $\textbf{A}^0$.  Combining Eqs.~(\ref{gauge1}) and~(\ref{repre_1}) we find

\begin{equation}\label{gauge_cell}
\nabla_y\cdot\textbf{a}_i = 0,
\end{equation}

\noindent which is a gauge condition for solving the cell problem in Eq.~(\ref{cell_1}).  Equations~(\ref{cell_1}) and~(\ref{gauge_cell}) must be solved in the unit cell of the metamaterial with periodic boundary conditions.  We note here that the corrector fields $\mathrm{a}_{0i}$ and $\textbf{a}_i$ have units of length.  Despite this, they can intuitively be thought of as the components of an electromagnetic four-potential.

Finally, the $\alpha^0$ order equation is

\begin{equation}\label{alpha_0}
\begin{array}{rl}
\left(\!\!\!\begin{array}{c}
\nabla_y\cdot \\ \nabla_y\times
\end{array}\!\!\!\right)\cdot\left[\hat{K}\cdot
\left(\!\!\!\begin{array}{c}
-\nabla_y\mathrm{A}_0^2-\nabla_x\mathrm{A}_0^1 \\
\nabla_y\times\textbf{A}^2+\nabla_x\times\textbf{A}^1
\end{array}\!\!\!\right)\right] + & \\ \\
\left(\!\!\!\begin{array}{c}
\nabla_x\cdot \\ \nabla_x\times
\end{array}\!\!\!\right)\cdot\left[\hat{K}\cdot
\left(\!\!\!\begin{array}{c}
-\nabla_x\mathrm{A}_0^0-\nabla_y\mathrm{A}_0^1 \\
\nabla_x\times\textbf{A}^0+\nabla_y\times\textbf{A}^1
\end{array}\!\!\!\right)\right] & = 0.
\end{array}
\end{equation}

\noindent Integrating this equation over the unit cell with respect to the $\textbf{y}$ variable causes the first term to vanish due to the periodic boundary conditions.  Using the representation for $\mathrm{A}_0^1$ and $\textbf{A}^1$ defined in Eq.~(\ref{repre_1}), the integrated Eq.~(\ref{alpha_0}) reduces to 

\begin{equation}\label{macro_field}
\left(\!\!\!\begin{array}{c}
\nabla_x\cdot \\ \nabla_x\times
\end{array}\!\!\!\right)\cdot\left[\bar{K}\cdot
\left(\!\!\!\begin{array}{c}
-\nabla_x \mathrm{A}_0 \\ \nabla_x\times\textbf{A}_0
\end{array}\!\!\!\right)\right] = 0,
\end{equation}

\noindent where the macroscopic constitutive matrix $\bar{K}$ is given by

\begin{equation}\label{macro_con}
\left(\bar{K}\right)_{ij} =  \displaystyle\frac{1}{V}\int_{\Omega} d^3y \ \ \hat{\textbf{e}}_i\cdot \hat{K} \cdot \left[\hat{\textbf{e}}_j + 
\left(\!\!\!\begin{array}{c}
-\nabla_y\mathrm{a}_{0j} \\ \nabla_y\times\textbf{a}_j
\end{array}\!\!\!\right)\right].
\end{equation}

\noindent Here $\int_{\Omega}d^3y$ indicates an integral over the unit cell and $V$ is the volume of the unit cell.  The inverse volume term insures that the units are correct and that the averaging procedure returns the correct result when averaging an already homogeneous material.

\subsection{Dual scalar-pseudoscalar homogenization}

As an alternative to using the electromagnetic four-potential to perform asymptotic homogenization, we can take advantage of the fact that we are operating in the static regime, allowing us to represent both the electric and magnetic fields as gradients of scalar and pseudoscalar fields respectively.  In addition to the previously used electrostatic scalar potential $\mathrm{A}_0$, we now introduce the magnetostatic pseudoscalar potential $\mathrm{C}_0$.  As before, we expand our fields in orders of the unitless small parameter $\alpha$

\begin{equation}\label{expan2}
\begin{array}{rl}
\mathrm{A}_0^{\alpha}(\textbf{x}) = & \mathrm{A}_0^0(\textbf{x},\textbf{y}) + \alpha \mathrm{A}_0^1(\textbf{x},\textbf{y}) + \alpha^2 \mathrm{A}_0^2(\textbf{x},\textbf{y}) + \cdots\ , \\[5pt]
\mathrm{C}_0^{\alpha}(\textbf{x}) = & \mathrm{C}_0^0(\textbf{x},\textbf{y}) + \alpha \mathrm{C}_0^1(\textbf{x},\textbf{y}) + \alpha^2 \mathrm{C}_0^2(\textbf{x},\textbf{y}) + \cdots\ .
\end{array}
\end{equation}

\noindent The electric and magnetic fields are defined as

\begin{equation}
\begin{array}{rl}
\textbf{E} = & -\nabla\mathrm{A}_0^{\alpha}, \\ \\
\textbf{H} = & -\nabla\mathrm{C}_0^{\alpha}.
\end{array}
\end{equation}

\noindent  Because the fields are static we can disregard any vector (or pseudovector) potential fields.  This method will produce the same static constitutive parameters as the method outlined in the previous section, but with lower computational cost.  Instead of solving for a single component scalar field and a three component vector field, while enforcing a gauge condition, as was done in the previous section, we are now solving for two single component scalar (pseudoscalar) fields without the need for a gauge condition.

The field equations that must be satisfied involve the divergences of the electric displacement and magnetic flux fields.

\begin{equation}\label{field2}
\begin{array}{rl}
0 = \nabla\cdot\textbf{D} = & \nabla\cdot\left[\hat{\epsilon}\cdot(-\nabla\mathrm{A}_0^{\alpha})+\hat{\xi}\cdot(-\nabla\mathrm{C}_0^{\alpha})\right], \\ \\
0 = \nabla\cdot\textbf{B} = & \nabla\cdot\left[\hat{\zeta}\cdot(-\nabla\mathrm{A}_0^{\alpha})+\hat{\mu}\cdot(-\nabla\mathrm{C}_0^{\alpha})\right].
\end{array}
\end{equation}

\noindent By putting the expanded fields Eq.~(\ref{expan2}) in to the field equation Eq.~(\ref{field2}), and separating terms with the same order of $\alpha$, we find a system of equations that must be satisfied.  The $\alpha^{-2}$ order equation is

\begin{equation}
\left(\!\!\!\begin{array}{c}
\nabla_y\cdot \\ \nabla_y\cdot\end{array}
\!\!\!\right)\cdot\left[\hat{C}\cdot
\left(\!\!\!\begin{array}{c}
-\nabla_y\mathrm{A}_0^0 \\[5pt] -\nabla_y\mathrm{C}_0^0
\end{array}\!\!\!\right)\right] = 0,
\end{equation}

\noindent where $\hat{C}$ is defined in Eq.~(\ref{Con2}).  This equation implies that the scalar (pseudoscalar) fields are independent of $\textbf{y}$ or $\mathrm{A}_0^0 = \mathrm{A}_0^0(\textbf{x})$ and $\mathrm{C}_0^0 = \mathrm{C}_0^0(\textbf{x})$.

The $\alpha^{-1}$ order equation is

\begin{equation}
\left(\!\!\!\begin{array}{c}
\nabla_y\cdot \\ \nabla_y\cdot\end{array}\!\!\!\right)
\cdot\left[\hat{C}\cdot
\left(\!\!\!\begin{array}{c}
-\nabla_x\mathrm{A}_0^0-\nabla_y\mathrm{A}_0^1 \\[5pt]
-\nabla_x\mathrm{C}_0^0-\nabla_y\mathrm{C}_0^1
\end{array}\!\!\!\right)\right] = 0.
\end{equation}

\noindent Here we have used the fact the both $0$th order fields are independent of $\textbf{y}$.  As is usual with the asymptotic homogenization method, we represent the first order fields in term of the derivatives of the $0$th order fields times correctors

\begin{equation}\label{repre_2}
\begin{array}{rl}
\mathrm{A}_0^1(\textbf{x},\textbf{y}) = & \displaystyle\sum_{i=1}^6
\left(\!\!\!\begin{array}{c}
-\nabla_x\mathrm{A}_0^0(\textbf{x}) \\ -\nabla_x\mathrm{C}_0^0(\textbf{x})
\end{array}\!\!\!\right)_i
\mathrm{a}_{0i}(\textbf{y}), \\[5pt]
\mathrm{C}_0^1(\textbf{x},\textbf{y}) = & \displaystyle\sum_{i=1}^6
\left(\!\!\!\begin{array}{c}
-\nabla_x\mathrm{A}_0^0(\textbf{x}) \\ -\nabla_x\mathrm{C}_0^0(\textbf{x})
\end{array}\!\!\!\right)_i
\mathrm{c}_{0i}(\textbf{y}).
\end{array}
\end{equation}

\noindent  This allows us to satisfy Eq.~(\ref{repre_2}) by solving the cell problem

\begin{equation}\label{cell_2}
\left(\!\!\!\begin{array}{c}
\nabla_y\cdot \\ \nabla_y\cdot
\end{array}\!\!\!\right)
\cdot\left[\hat{C}\cdot\left(\hat{\textbf{e}}_i + 
\left(\!\!\!\begin{array}{c}
-\nabla_y\mathrm{a}_{0i} \\ -\nabla_y\mathrm{c}_{0i}
\end{array}\!\!\!\right)
\right)\right] = 0,
\end{equation}

\noindent remembering that the correctors $\mathrm{a}_{0i}(\textbf{y})$ and $\mathrm{c}_{0i}(\textbf{y})$ are periodic with respect to $\textbf{y}$.  Again, we note that the corrector fields $\mathrm{a}_{0i}$ and $\mathrm{c}_{0i}$ have units of length, but can be thought of as electrostatic and magnetostatic potentials respectively.

The $\alpha^0$ field equation is

\begin{equation}\label{scalar_alpha_0}
\begin{array}{rl}
\left(\!\!\!\begin{array}{c}
\nabla_y\cdot \\ \nabla_y\cdot
\end{array}\!\!\!\right)
\cdot\left[\hat{C}\cdot
\left(\!\!\!\begin{array}{c}
-\nabla_y\mathrm{A}_0^2-\nabla_x\mathrm{A}_0^1 \\
-\nabla_y\mathrm{C}_0^2-\nabla_x\mathrm{C}_0^1
\end{array}\!\!\!\right)\right] + & \\ \\
\left(\!\!\!\begin{array}{c}
\nabla_x\cdot \\ \nabla_x\cdot
\end{array}\!\!\!\right)
\cdot\left[\hat{C}\cdot
\left(\!\!\!\begin{array}{c}
-\nabla_x\mathrm{A}_0^0-\nabla_y\mathrm{A}_0^1 \\
-\nabla_x\mathrm{C}_0^0-\nabla_y\mathrm{C}_0^1
\end{array}\!\!\!\right)\right] & = 0.
\end{array}
\end{equation}

\noindent Integrating this equation over the unit cell with respect to the variable $\textbf{y}$ causes the first term to vanish due to the periodicity of the fields.  Using the representation provided in Eq.~(\ref{repre_2}), the integrated Eq.~(\ref{scalar_alpha_0}) becomes the macroscopic field equation

\begin{equation}
\left(\!\!\!\begin{array}{c}
\nabla_x\cdot \\ \nabla_x\cdot\end{array}\!\!\!\right)
\cdot\left[\bar{C}\cdot
\left(\!\!\!\begin{array}{c}
-\nabla_x\mathrm{A}_0^0 \\
-\nabla_x\mathrm{C}_0^0
\end{array}\!\!\!\right)\right] = 0,
\end{equation}

\noindent where the macroscopic constitutive matrix $\bar{C}$ is given by

\begin{equation}\label{macro_con_2}
\left(\bar{C}\right)_{ij} = \displaystyle\frac{1}{V}\int_{\Omega}d^3y \ \ \hat{\textbf{e}}_i\cdot\hat{C}\cdot\left[\hat{\textbf{e}}_j + 
\left(\!\!\!\begin{array}{c}
-\nabla_y\mathrm{a}_{0j} \\
-\nabla_y\mathrm{c}_{0j}
\end{array}\!\!\!\right)\right].
\end{equation}

We note that the cell problem Eq.~(\ref{cell_2}) and the formula for the macroscopic constitutive parameters Eq.~(\ref{macro_con_2}) are equivalent to those in Ref. \onlinecite{Ouchetto_06}.  However, our derivation follows the standard asymptotic expansion method and we have identified $\mathrm{C}_0$ as the magnetostatic pseudoscalar potential.  Our homogenization method uses the potential fields whereas Ref. \onlinecite{Ouchetto_06} attempts to homogenize the electric and magnetic fields, and introduces two scalar fields without physical justification.

Finally, a third formulation of the electro-magnetostatic asymptotic homogenization procedure is possible using vector and pseudovector potentials.  The electric displacement and magnetic flux density would be defined as 

\begin{equation}
\begin{array}{rl}
\textbf{D} = & -\nabla\times\textbf{C}, \\
\textbf{B} = & \nabla\times\textbf{A}.
\end{array}
\end{equation}

\noindent Here $\textbf{A}$ is the electromagnetic vector potential used in Sec.\ref{Sec_2}A, and $\textbf{C}$ is a magnetoelectric pseudovector potential, the dual field of the electromagnetic vector potential.  This method should return the same homogenization results as the two previously described methods, but would be computationally expensive.  It would require solving for two different 3 component vector (pseudovector) fields while enforcing two different gauge conditions.  Though it is hard to imagine a situation where this method would be advantageous, it is still interesting to note that it exists as an option.

\section{One dimensinoal layered chiral metamaterial}\label{Sec_3}

As a first example of this long wavelength homogenization procedure, we examine a one dimensional metamaterial consisting of periodically layered materials, some of these layers being microscopically chiral.

\begin{figure}[h]
\begin{center}
\includegraphics[width=0.5\columnwidth]{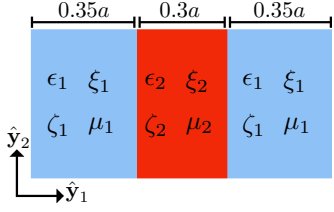}
\end{center}
\caption{Unit cell of a layered metamaterial.  The metamaterial is periodic in the $\hat{\textbf{y}}_1$ direction with periodic lattice constant $a$.  Translations in the $\hat{\textbf{y}}_2$ and $\hat{\textbf{y}}_3$ directions leave the geometry unchanged.  The unit cell consists of two different layers, each with the constitutive parameters shown in Eq.~(\ref{one_D_con}).}\label{Fig_1}
\end{figure}

Fig.~\ref{Fig_1} shows the unit cell of the layered metamaterial.  The isotropic constitutive parameters of the two different layers are

\begin{equation}\label{one_D_con}
\begin{array}{rlcrl}
\epsilon_1= & 1, &\ \ \ \ & \epsilon_2= & \ \ \ \! 5, \\[5pt]
\xi_1= & 0, &\ \ \ \ & \xi_2= & \ \ \ \! 2.85\mathrm{i}, \\[5pt]
\zeta_1= & 0, &\ \ \ \ & \zeta_2= & -2.85\mathrm{i}, \\[5pt]
\mu_1= & 1, &\ \ \ \ & \mu_2= & \ \ \ \! 1.
\end{array}
\end{equation}

\noindent The subscripts $1$ and $2$ indicate that the constitutive parameters correspond to layers $1$ and $2$ (see Fig.~\ref{Fig_1}).  Notice that the microscopic constitutive parameters we are using are reciprocal ($\hat{\epsilon}=\hat{\epsilon}^{\mathsf{T}}$, $\hat{\mu}=\hat{\mu}^{\mathsf{T}}$ and $\hat{\xi}=-\hat{\zeta}^{\mathsf{T}}$ where $^\mathsf{T}$ indicates the transpose) though this assumption is not necessary for asymptotic homogenization.  Also, for simplicity we have made the chirality of the second layer independent of frequency, despite the fact that physically a lossless chirality should be proportional to $\omega$ in the long wavelength limit.

\begin{figure*}[t]
\begin{center}
\includegraphics[width=\textwidth]{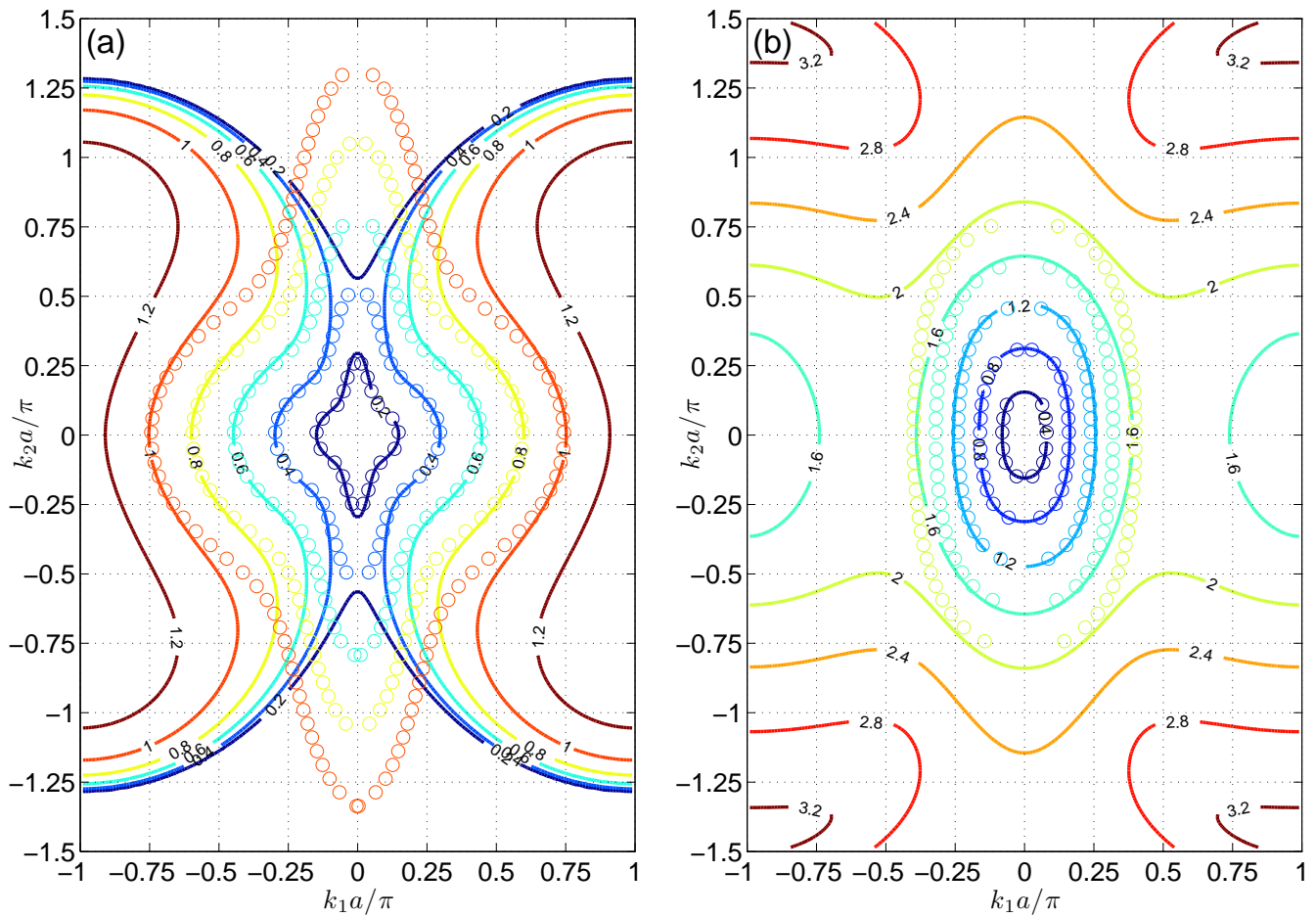}
\end{center}
\caption{Isofrequency contours for (a) left handed and (b) right handed elliptically polarized waves calculated with an eigenvalue simulation of the layered structure shown in Fig.~\ref{Fig_1} (solid lines) and from the dispersion relation for a homogeneous medium with the constitutive parameters of Eq.~(\ref{one_D_macro_con}) (circles).  The labels on the solid lines are the normalized frequency $\omega a/c$.  The $\omega a/c=1.6$ lobes on the left and right hand sides of Fig.~\ref{Fig_2}(b) are left handed modes in the second propagating band.}\label{Fig_2}
\end{figure*}

In the appendix of this paper we use the asymptotic homogenization procedure outlined in Sec.~\ref{four_po} to analytically solve for the macroscopic constitutive parameters of a one dimensional layered structure.  The resulting macroscopic constitutive parameters are

\begin{widetext}
\begin{equation}\label{C_sol}
\begin{array}{c}
\bar{C} = 
\left(\!\!\!\begin{array}{cccccc}
\epsilon_{\perp} & 0 & 0 & \xi_{\perp} & 0 & 0 \\
0 & \epsilon_{\parallel} & 0 & 0 & \xi_{\parallel} & 0\\
0 & 0 & \epsilon_{\parallel} & 0 & 0 & \xi_{\parallel} \\
\zeta_{\perp} & 0 & 0 & \mu_{\perp} & 0 & 0 \\
0 & \zeta_{\parallel} & 0 & 0 & \mu_{\parallel} & 0 \\
0 & 0 & \zeta_{\parallel} & 0 & 0 & \mu_{\parallel}
\end{array}\!\!\!\right), \\ \\
\begin{array}{rlcrl}
\epsilon_{\perp} = & \displaystyle \frac{\langle\epsilon/(\epsilon\mu-\xi\zeta)\rangle}{\langle \epsilon/(\epsilon\mu-\xi\zeta)\rangle \langle\mu/(\epsilon\mu-\xi\zeta)\rangle - \langle\xi/(\epsilon\mu-\xi\zeta)\rangle \langle\epsilon/(\epsilon\mu-\xi\zeta)\rangle},
& \ \ \ \ \ & \epsilon_{\parallel} = & \langle\epsilon\rangle, \\ \\
\xi_{\perp} = & \displaystyle \frac{\langle\xi/(\epsilon\mu-\xi\zeta)\rangle}{\langle \epsilon/(\epsilon\mu-\xi\zeta)\rangle \langle\mu/(\epsilon\mu-\xi\zeta)\rangle - \langle\xi/(\epsilon\mu-\xi\zeta)\rangle \langle\xi/(\epsilon\mu-\xi\zeta)\rangle},
& \ \ \ \ \ & \xi_{\parallel} = & \langle\xi\rangle, \\ \\
\zeta_{\perp} = & \displaystyle \frac{\langle\zeta/(\epsilon\mu-\xi\zeta)\rangle}{\langle \epsilon/(\epsilon\mu-\xi\zeta)\rangle \langle\mu/(\epsilon\mu-\xi\zeta)\rangle - \langle\xi/(\epsilon\mu-\xi\zeta)\rangle \langle\zeta/(\epsilon\mu-\xi\zeta)\rangle},
& \ \ \ \ \ & \zeta_{\parallel} = & \langle\zeta\rangle, \\ \\
\mu_{\perp} = & \displaystyle \frac{\langle\mu/(\epsilon\mu-\xi\zeta)\rangle}{\langle \epsilon/(\epsilon\mu-\xi\zeta)\rangle \langle\mu/(\epsilon\mu-\xi\zeta)\rangle - \langle\xi/(\epsilon\mu-\xi\zeta)\rangle \langle\mu/(\epsilon\mu-\xi\zeta)\rangle},
& \ \ \ \ \ & \mu_{\parallel} = & \langle\mu\rangle.
\end{array}
\end{array}
\end{equation}
\end{widetext}

\noindent Here $\langle X \rangle$ indicates the arithmetic mean of $X$ over the unit cell

\begin{equation}\label{average_eq}
\langle X \rangle = \displaystyle\frac{1}{a}\int_{0}^{a}\!\!\!\!dy_1\ X(y_1),
\end{equation}

\noindent where $a$ is the lattice constant of the unit cell.  In addition to the crystal being one dimensional, the only other assumption made in the derivation is that the microscopic constitutive parameters are bi-isotropic.

After applying the formulas for the macroscopic constitutive parameters to the layered structure in Fig.~\ref{Fig_1} we calculate

\begin{equation}\label{one_D_macro_con}
\begin{array}{rlcrl}
\epsilon_{\perp} = & \ \ \ 3.81, &\ \ \ \ & \epsilon_{\parallel} = & \ \ \ \ \! 2.20, \\[5pt]
\xi_{\perp} = & -\mathrm{i}4.75, &\ \ \ \ & \xi_{\parallel} = & \ \ \ \! \mathrm{i}0.855, \\[5pt]
\zeta_{\perp} = & \ \ \ \! \mathrm{i}4.75, &\ \ \ \ & \zeta_{\parallel} = & -\mathrm{i}0.855, \\[5pt]
\mu_{\perp} = & \ \ 10.5, &\ \ \ \ & \mu_{\parallel} = & \ \ \ \ \! 1.
\end{array}
\end{equation}

\noindent Here $_\perp$ indicates the direction perpendicular to the layer interfaces or $\hat{\textbf{y}}_1$, and $_\parallel$ indicates the directions parallel to the layer interfaces or $\hat{\textbf{y}}_2$ and $\hat{\textbf{y}}_3$.

Looking at the long wavelength constitutive parameters in Eq.~(\ref{one_D_macro_con}), we see that $\mu_{\perp}$ is non-unity, despite the fact that no microscopically magnetic materials were present in the unit cell.  This is due to the presence of the chiral layer.  It is possible to create a magnetic response by adding bi-isotropic materials to a metamaterial, though this comes at the cost of making the macroscopic material bi-anisotropic.  Second, notice that $\xi_{\perp}$ has a sign opposite that of $\xi_{\parallel}$, the same being true for different components of $\zeta$ as well.  Finally, the macroscopic constitutive parameters are reciprocal, which is to be expected due to the fact that the microscopic constitutive parameters are reciprocal as well.

To test these constitutive parameters we compare the dispersion relation of waves in a homogeneous effective medium with the constitutive parameters in Eq.~(\ref{one_D_macro_con}), to the dispersion relation of waves in the inhomogeneous layered medium.  Fig.~\ref{Fig_2} plots an isofrequency contour~\cite{Joannopoulos_08,Sakoda_04} of the layered metamaterial calculated using a finite element eigenvalue simulation (solid lines) as well as isofrequency contours of a homogeneous medium with the constitutive parameters shown in Eq.~(\ref{one_D_macro_con}) (circles).  Fig.~\ref{Fig_2} plots two modes, which are left and right handed elliptically polarized.  The isofrequency contours for the homogeneous effective medium were calculated by solving the Maxwell equations in $\omega$ and $\textbf{k}$ space as a $6\times 6$ eigenvalue problem.  The eigenvectors correspond to the electric and magnetic fields and were used to determine the handedness of each mode.  This determination was done using the convention that a left handed elliptically polarized mode has an electric field at a fixed point moving counterclockwise around the wave-vector in time as seen from the point of view of someone the wave-vector is pointing away from.

We see from Fig.~\ref{Fig_2} that the two dispersion relations agree very well when $\omega a/c$, $k_1a$ and $k_2a$ are all small.  For large frequencies or wavenumbers, the dispersion relations diverge from each other.  This is to be expected.  The asymptotic homogenization method is only intended to be used in the long wavelength limit.  For large frequencies unit cell inclusions can become resonant and at large wavenumbers spatial dispersion can become significant, both causing the long wavelength constitutive parameters to fail.

\section{Three dimensional uniaxial bianisotropic metamaterial}

\begin{figure}[h]
\begin{center}
\includegraphics[width=0.5\columnwidth]{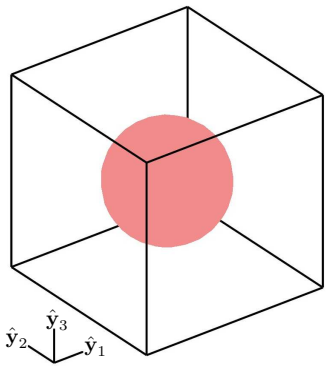}
\end{center}
\caption{Unit cell of a three dimensional bianisotropic metamaterial.  The crystal has a cubic lattice with lattice constant $a$.  Centered in the unit cell is a sphere of radius $0.3a$ consisting of an isotropic Tellegen material with constitutive parameters given in the text.  The material outside the sphere is a uniaxial dielectric with permittitivities provided in the text.}\label{Fig_3}
\end{figure}

As a second example of bianisotropic asymptotic homogenization, we consider a three dimensional crystal pictured in Fig.~\ref{Fig_3}.

\noindent  The unit cell consists of cube with lattice constant $a$ with a sphere at the center with radius $0.3a$.  The sphere is a nonreciprocal isotropic Tellegen material with the constitutive parameters

\begin{equation}
\begin{array}{rlcrl}
\epsilon = & 1.78, &\ \ \ \ & \xi = & 2, \\[5pt]
\zeta = & 2, &\ \ \ \ & \mu = & 1.
\end{array}
\end{equation}

\noindent The material surrounding the sphere is a reciprocal uniaxial dielectric with the constitutive parameters
\begin{equation}
\begin{array}{rlcrl}
\epsilon_{\perp} = & 4, &\ \ \ \ & \epsilon_{\parallel} = & 1.5, \\[5pt]
\xi = \zeta = & 0, &\ \ \ \ & \mu = & 1,
\end{array}
\end{equation}

\noindent where $\perp$ indicates the $\hat{\textbf{y}}_1$ direction and $\parallel$ indicates the $\hat{\textbf{y}}_2$ and $\hat{\textbf{y}}_3$ directions.

The crystal shown in Fig.~\ref{Fig_3} was homogenized with both of the asymptotic methods presented in Sec.~\ref{Sec_2} using the commercial finite element software Comsol Multiphysics 3.5a.  Copies of these homogenization simulations are available upon request by contacting the author using the email address preceding the references.  The   Upon applying either of the bianisotropic asymptotic homogenization procedures outlined earlier in this paper, the macroscopic constitutive parameters are found to be

\begin{equation}\label{3_D_macro_con}
\begin{array}{rlcrl}
\epsilon_{\perp} = & 3.43, &\ \ \ \ & \epsilon_{\parallel} = & 1.38, \\[5pt]
\xi_{\perp} = & 0.335, &\ \ \ \ & \xi_{\parallel} = & 0.267, \\[5pt]
\zeta_{\perp} = & 0.335, &\ \ \ \ & \zeta_{\parallel} = & 0.267, \\[5pt]
\mu_{\perp} = & 0.936, &\ \ \ \ & \mu_{\parallel} = & 0.910.
\end{array}
\end{equation}

We can see that although the geometry of the unit cell is very symmetric, with several reflection and rotational symmetries, the macroscopic constitutive parameters are uniaxial and bianisotropic.  This is because the microscopic constitutive parameters break the geometric symmetries of the unit cell.  The uniaxial dielectric surrounding the sphere breaks several of the rotational symmetries of the unit cell, and the sphere consisting of the Tellegen material breaks all of the reflection symmetries.

\begin{figure*}[t]
\begin{center}
\includegraphics[width=\textwidth]{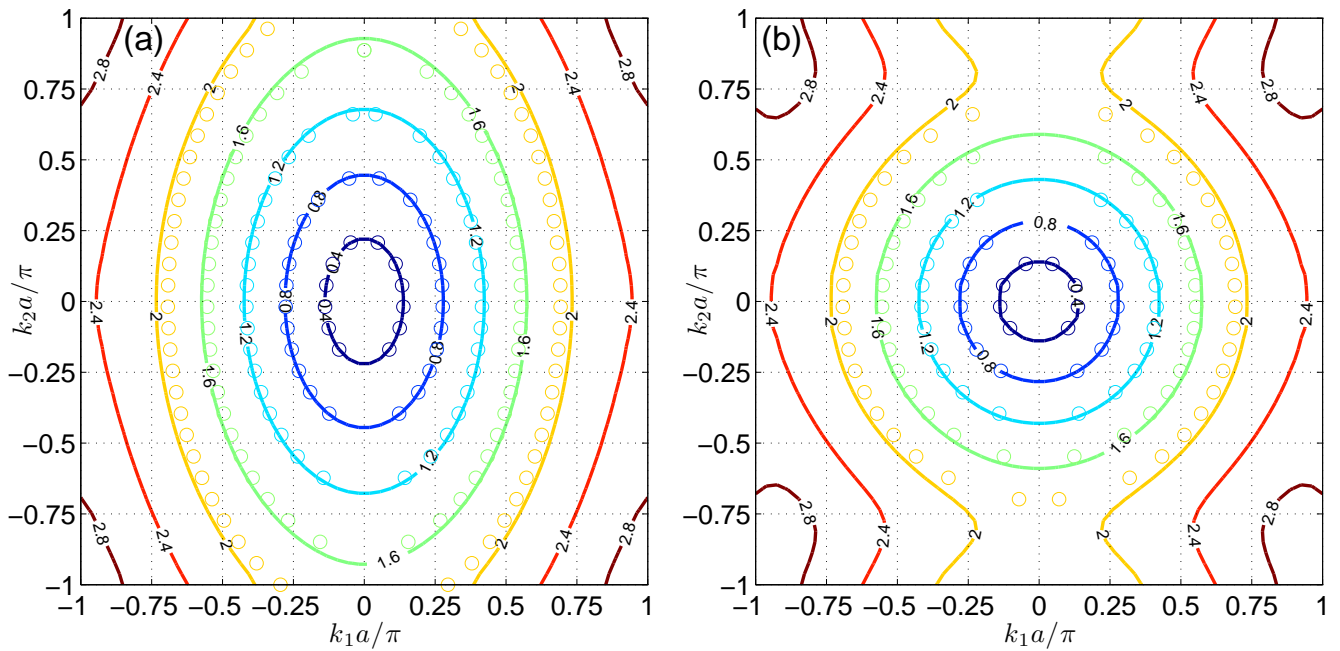}
\end{center}
\caption{Isofrequency contours for the three dimensional crystal shown in Fig.~\ref{Fig_3}.  Both isofrequency contours (a) and (b) are for different linearly polarized eigenmodes.  The isofrequency contours are calculated with a eigenvalue simulation of the crystal in Fig.~\ref{Fig_3} (solid lines) and from the dispersion relation for a homogeneous medium with the macroscopic constitutive parameters given in Eq.~(\ref{3_D_macro_con}) (circles).}\label{Fig_4}
\end{figure*}

We test these macroscopic constitutive parameters by using them to calculate the dispersion relation of the crystal in Fig.~\ref{Fig_3}.  Fig.~\ref{Fig_4} shows two isofrequency contour diagrams, one calculated with a finite element eigenvalue simulation of the crystal in Fig.~\ref{Fig_3} (solid lines), and one calculated with the long wavelength constitutive parameters calculated with the asymptotic homogenization theory.  From Fig.~\ref{Fig_4}, we can clearly see that the two isofrequency contours agree in the long wavelength limit, but diverge for larger frequencies and wavenumbers.  The dispersion relation calculated from the macroscopic constitutive parameters was obtained by solving the Maxwell equations in $\omega$ and $\textbf{k}$ space as a $6\times 6$ eigenvalue problem.  The resulting eigenvectors represent the macroscopic electric and magnetic fields.  These eigenvectors have the same phase for the different components of the electric and magnetic fields, indicating that the eigenmodes are linearly polarized.  Despite this, the eigenmode polarizations do not in general lie upon the principle axes of the crystal.

As a final note, looking at Fig.~\ref{Fig_4}, one can see that the contours for the eigenmode on the left side of Fig~\ref{Fig_4} seem to continuously merge with the contours of the eigenmode on the right side of Fig.~\ref{Fig_4} across the $k_1a/\pi=\pm1$ boundaries.  A similar relationship exists at the $k_2a/\pi=\pm1$ boundaries.  If one takes the two isofrequency contours in Fig.~\ref{Fig_4}, and tiles them in a checkerboard fashion, one sees that the contours for one eigenmode continuously merge with the contours of the other eigenmode as they cross the boundaries of the Brillouin zone.  It is not clear why this happens, but it only seems to occur with a nonreciprocal bianisotropic crystal.

\section{Conclusion}
In conclusion, we have presented two different implementations of the asymptotic homogenization method for electromagnetic metamaterials with bianisotropic inclusions, including one method that handles a gauge condition.  Because they are asymptotic methods, they are only valid in the long wavelength limit.  We have tested the homogenization methods by comparing dispersion relations calculated from the resulting constitutive parameters with the true dispersion relations returned by eigenvalue simulations and found good agreement in the long wavelength limit.  It is hoped that in the future these homogenization methods, particularly the four-potential method, can be generalized beyond the long wavelength limit to describe temporal and spatial dispersion effects.

\section*{Acknowledgements}
Chris Fietz would like to acknowledge support from the IC Postdoctoral Research Fellowship Program.

\appendix
\section{Analytic homogenization of layered structure}\label{Appendix_1}

Here we analytically solve for the long wavelength constitutive parameters of a one dimensional layered structure.  We do so using the four-potential method presented in Sec.~\ref{four_po}.  The layered structure will have constitutive parameters that only vary in the $y_1$ direction.  They are constant in the $y_2$ and $y_3$ directions.  This simplifies the problem considerably, allowing us to obtain an analytic solution.  We assume that the constitutive parameters are bi-isotropic, thus all four tensors (pseudotensors) are symmetric under a spatial rotation and thus can be represented as scalars (pseudoscalars)

\begin{equation}
\hat{K} = 
\left(\!\!\!\begin{array}{cccccc}
p & 0 & 0 & l & 0 & 0 \\
0 & p & 0 & 0 & l & 0 \\
0 & 0 & p & 0 & 0 & l \\
m & 0 & 0 & q & 0 & 0 \\
0 & m & 0 & 0 & q & 0 \\
0 & 0 & m & 0 & 0 & q
\end{array}\!\!\!\right).
\end{equation}

The first step is to solve the cell problem in Eq.~(\ref{cell_1}).  To do this we first use the symmetry of the unit cell to simplify the cell problem.  Due to the translational symmetry of the unit cell in the $y_2$ and $y_3$ directions, the microscopic fields only depend on $y_1$, or $\mathrm{a}_{0i}(\textbf{y})=\mathrm{a}_{0i}(y_1)$ and $\textbf{a}_i(\textbf{y})=\textbf{a}_i(y_1)$.  Any derivatives with respect to $y_2$ or $y_3$ vanish in both the cell problem Eq.~(\ref{cell_1}) and the cell gauge condition Eq.~(\ref{gauge_cell}).  The cell gauge condition becomes $\partial\mathrm{a}_{1i}/\partial y_1 = 0$.  Thus all $\partial\mathrm{a}_{1i}/\partial y_1$ terms in Eq.~(\ref{cell_1}) vanish.  This leaves us with three out of the original four equations

\begin{equation}\label{cell_one}
\begin{array}{rl}
\displaystyle\frac{\partial}{\partial y_1}\left[p\left(-\frac{\partial\mathrm{a}_{0i}}{\partial y_1}+\delta_{1i}\right)+l\delta_{4i}\right] = & 0, \\ \\
\displaystyle\frac{\partial}{\partial y_1}\left[m\delta_{2i}+q\left(-\frac{\partial\mathrm{a}_{3i}}{\partial y_1}+\delta_{5i}\right)\right] = & 0 \\ \\
\displaystyle-\frac{\partial}{\partial y_1}\left[m\delta_{3i}+q\left(\frac{\mathrm{a}_{2i}}{\partial y_1}+\delta_{6i}\right)\right] = & 0.
\end{array}
\end{equation}

Equation~(\ref{cell_one}) contains three different equations that must be solved with six different values for the index $i$, yielding 18 different equations overall.  These equations must be solved with periodic boundary conditions on the $y_1$ domain.  It should be noted that these three differential equations are decoupled from each other and can be solved separately.  Second, all three equations are for fields depending on a single dimension ($y_1$).  Finally, all three fall into one of three types of ordinary differential equation problems

\begin{equation}
\begin{array}{rl}
\displaystyle\frac{\partial}{\partial y_1}\left[\mathrm{C}_1\left(\frac{\partial\psi}{\partial y_1}+1\right)\right]=&0, \\ \\
\displaystyle\frac{\partial}{\partial y_1}\left[\mathrm{C}_2+\mathrm{C}_3\frac{\partial\psi}{\partial y_1}\right]=&0, \\ \\
\displaystyle\frac{\partial}{\partial y_1}\left[\mathrm{C}_4\frac{\partial\psi}{\partial y_1}\right]=&0,
\end{array}
\end{equation}

\begin{table}[h]
\centering
\begin{tabular}{|c|c|c|c|}
\cline{2-4}
\multicolumn{1}{c|}{}\parbox[0pt][30pt][c]{10pt}{}& $\displaystyle-\frac{\partial\mathrm{a}_{0i}}{\partial y_1}$ & $\displaystyle-\frac{\partial\mathrm{a}_{3i}}{\partial y_1}$ & $\displaystyle\frac{\partial\mathrm{a}_{2i}}{\partial y_1}$ \\[10pt] \cline{1-4}
\parbox[0pt][30pt][c]{10pt}{}$i = 1$ & $\displaystyle\frac{1}{p\langle1/p\rangle}-1$ & $0$ & $0$ \\[10pt]\cline{1-4}
\parbox[0pt][30pt][c]{10pt}{}$i = 2$ & $0$ & $\displaystyle\frac{1}{q}\frac{\langle m/q\rangle}{\langle 1/q\rangle}-\frac{m}{q}$ & $0$ \\[10pt]\cline{1-4}
\parbox[0pt][30pt][c]{10pt}{}$i = 3$ & $0$ & $0$ & $\displaystyle\frac{1}{q}\frac{\langle m/q\rangle}{\langle 1/q\rangle}-\frac{m}{q}$ \\[10pt]\cline{1-4}
\parbox[0pt][30pt][c]{10pt}{}$i = 4$ & $\displaystyle\frac{1}{p}\frac{\langle l/p\rangle}{\langle 1/p\rangle}-\frac{l}{p}$ & $0$ & $0$ \\[10pt]\cline{1-4}
\parbox[0pt][30pt][c]{10pt}{}$i = 5$ & $0$ & $\displaystyle\frac{1}{q\langle1/q\rangle}-1$ & $0$ \\[10pt]\cline{1-4}
\parbox[0pt][30pt][c]{10pt}{}$i = 6$ & $0$ & $0$ & $\displaystyle\frac{1}{q\langle1/q\rangle}-1$ \\[10pt]\cline{1-4}
\end{tabular}
\caption{Solutions to Eq.~(\ref{cell_one}).}
\label{cell_sol_table}
\end{table}

\noindent all solved with periodic boundary conditions on $y_1$.  Here $\psi$ is a scalar field that only depends on $y_1$ and $\mathrm{C}_{1-4}$ are functions of $y_1$ representing the various constitutive parameters in Eq.~(\ref{cell_one}).  The solutions to these ordinary differential equations are easily found and are shown in Table~\ref{cell_sol_table}.  Here the averaging operation $\langle \rangle$ is defined in Eq.~(\ref{average_eq}).

Now that we have the solutions to the cell problem we can calculate the macroscopic constitutive parameters.  Using the simplifications mentioned above due to the one dimensionality of the unit cell, Eq.~(\ref{macro_con}) becomes

\begin{equation}\label{K_sol}
(\bar{K})_{ij} = \displaystyle\frac{1}{a}\int_0^{a}\!\!\!\! dy_1\ \ \hat{\textbf{e}}_i\cdot\hat{K}\cdot
\left(\!\!\!\begin{array}{c}
\displaystyle-\frac{\partial\mathrm{a}_{0j}}{\partial y_1}+\delta_{1j} \\[5pt]
\delta_{2j} \\[5pt]
\delta_{3j} \\[5pt]
\delta_{4j} \\[5pt]
\displaystyle-\frac{\partial\mathrm{a}_{3j}}{\partial y_1}+\delta_{5j} \\[5pt]
\displaystyle\frac{\partial\mathrm{a}_{2j}}{\partial y_1}+\delta_{6j}
\end{array}\!\!\!\right),
\end{equation}

\noindent which when evaluated with the solutions to the cell problem in Table~\ref{cell_sol_table}. yields

\begin{equation}
\begin{array}{c}
\bar{K} = 
\left(\!\!\!\begin{array}{cccccc}
p_{\perp} & 0 & 0 & l_{\perp} & 0 & 0 \\
0 & p_{\parallel} & 0 & 0 & l_{\parallel} & 0\\
0 & 0 & p_{\parallel} & 0 & 0 & l_{\parallel} \\
m_{\perp} & 0 & 0 & q_{\perp} & 0 & 0 \\
0 & m_{\parallel} & 0 & 0 & q_{\parallel} & 0 \\
0 & 0 & m_{\parallel} & 0 & 0 & q_{\parallel}
\end{array}\!\!\!\right), \\ \\
\begin{array}{rlcrl}
p_{\perp} = & \displaystyle\frac{1}{\langle 1/p\rangle}, & \ \ \ \ &
p_{\parallel} = & \displaystyle\left\langle p-\frac{lm}{q}\right\rangle+\frac{\langle l/q\rangle\langle m/q\rangle}{\langle 1/q\rangle}, \\[12pt]
\end{array} \\[12pt]
\begin{array}{rlcrl}
l_{\perp} = & \displaystyle\frac{\langle l/p\rangle}{\langle 1/p\rangle}, & \ \ \ \ &
l_{\parallel} = & \displaystyle\frac{\langle l/q\rangle}{\langle 1/q\rangle}, \\[12pt]
m_{\perp} = & \displaystyle\frac{\langle m/p\rangle}{\langle 1/p\rangle}, & \ \ \ \ &
m_{\parallel} = & \displaystyle\frac{\langle m/q\rangle}{\langle 1/q\rangle}, \\[12pt]
\end{array} \\[12pt]
\begin{array}{rlcrl}
q_{\perp} = & \displaystyle\left\langle q-\frac{lm}{p}\right\rangle+\frac{\langle l/p\rangle\langle m/p\rangle}{\langle 1/p\rangle}, & \ \ \ \ &
q_{\parallel} = & \displaystyle\frac{1}{\langle 1/q\rangle},
\end{array}
\end{array}
\end{equation}

\noindent Converting the $\bar{K}$ constitutive parameters in Eq.~(\ref{K_sol}) back into the $\bar{C}$ constitutive parameters gives us Eq.~(\ref{C_sol}).  It is easy to see that in the absence of the bi-isotropic parameters ($\xi=\zeta=0$) the macroscopic constitutive parameters reduce to the standard expressions.

\end{document}